\documentclass[prl,reprint,superscriptaddress]{revtex4-1}

\usepackage{graphicx}  %
\usepackage{bm}        %
\usepackage{amssymb}   %

 \usepackage{color}

  \usepackage{ulem}

\begin{document}

\title{Ultrafast modification of Hubbard $U$ in a strongly correlated material: \\
       \textit{ab initio} high-harmonic generation in NiO}

 \author{Nicolas Tancogne-Dejean}
  \email{nicolas.tancogne-dejean@mpsd.mpg.de}
  \affiliation{Max Planck Institute for the Structure and Dynamics of Matter and Center for Free-Electron Laser Science, Luruper Chaussee 149, 22761 Hamburg, Germany}
 \affiliation{European Theoretical Spectroscopy Facility (ETSF)}

  \author{Michael A. Sentef}
  \affiliation{Max Planck Institute for the Structure and Dynamics of Matter and Center for Free-Electron Laser Science, Luruper Chaussee 149, 22761 Hamburg, Germany}
 
 \author{Angel Rubio}
  \email{angel.rubio@mpsd.mpg.de}
\affiliation{Max Planck Institute for the Structure and Dynamics of Matter and Center for Free-Electron Laser Science, Luruper Chaussee 149, 22761 Hamburg, Germany}
 \affiliation{European Theoretical Spectroscopy Facility (ETSF)}
 \affiliation{Nano-Bio Spectroscopy Group, Universidad del Pa\'is Vasco, , 20018 San Sebasti\'an, Spain }
\affiliation{Center for Computational Quantum Physics (CCQ), The Flatiron Institute, 162 Fifth Avenue, New York NY 10010}

\begin{abstract}
Engineering effective electronic parameters is a major focus in condensed matter physics. Their dynamical modulation opens the possibility of creating and controlling physical properties in systems driven out of equilibrium.
In this work, we demonstrate that the Hubbard $U$, the on-site Coulomb repulsion in strongly correlated materials, can be modified on femtosecond time scales by a strong nonresonant laser excitation in the prototypical charge transfer insulator NiO. Using our recently developed time-dependent density functional theory plus self-consistent $U$ (TDDFT+U) method, we demonstrate the importance of a dynamically modulated $U$ in the description of the high-harmonic generation of NiO. Our study opens the door to novel ways of modifying effective interactions in strongly correlated materials via laser driving, which may lead to new control paradigms for field-induced phase transitions and perhaps laser-induced Mott insulation in charge-transfer materials.
\end{abstract}
     
\maketitle

The interaction of strong subresonant driving fields with solids just below their damage threshold leads to very interesting phenomena, such as high-harmonic generation (HHG)~\cite{ghimire_observation_2011, schubert_sub-cycle_2014, hohenleutner_real-time_2015, luu_extreme_2015,ndabashimiye_solid-state_2016,vampa_linking_2015,liu_high-harmonic_2016,you_anisotropic_2016,langer_symmetry-controlled_2017}.
HHG has, so far, only been interpreted in a pure single-particle band-structure picture, assuming that electrons behave as independent particles during the interaction with the laser pulse. In this simplified view, only the electronic occupations of the bands change and the band structure of a solid remains frozen during the light-matter interaction, allowing for an interpretation in terms of interband and intraband mechanisms within the single-particle band structure of the solid. 

However, we know that laser driving can modify both the band structure and its topology (\cite{oka_photovoltaic_2009, lindner_floquet_2011, wang_observation_2013, mahmood_selective_2016, sentef_theory_2015, hubener_creating_2017}) and the effective interactions (\cite{raines_enhancement_2015, kennes_transient_2017, sentef_light-enhanced_2017, knap_dynamical_2016, komnik_bcs_2016, babadi_theory_2017, murakami_nonequilibrium_2017, mazza_nonequilibrium_2017, pomarico_enhanced_2017,  coulthard_enhancement_2017}) in solids, rendering the frozen-band and frozen-interaction scenarios questionable in many cases, in particular in the context of light-induced metal-insulator transitions \cite{rini_control_2007, caviglia_ultrafast_2012, stojchevska_ultrafast_2014}, magnetic systems \cite{kimel_nonthermal_2007, kirilyuk_ultrafast_2010, forst_driving_2011, forst_spatially_2015, mentink_ultrafast_2015, nova_effective_2017}, and superconductivity \cite{fausti_light-induced_2011, kaiser_optically_2014, hu_optically_2014, mitrano_possible_2016} among others.

The usual argument for neglecting dynamical changes of the electron-electron interaction in strong-field physics is that the strong excitation induced by the intense driving field completely dominates over these effects ~\cite{huttner_ultrahigh_2017}. 
For atoms in strong fields, the correlation effects are usually small, even if they can be relevant in specific cases~\cite{shiner_probing_2011}.

In solids, it was recently shown that assuming independent electrons is a perfectly valid approximation for HHG from bulk silicon~\cite{tancogne-dejean_impact_2017}, and that it performs also very well for bulk MgO~\cite{tancogne-dejean_ellipticity_2017}.
However, 
the general validity of the frozen band-structure approximation in strong driving fields must be questioned in correlated materials such as transition metal oxides.
Already, it was shown using time-dependent Hartree-Fock calculations that excitonic effects modify the HHG spectrum of a one-dimensional chain of H atoms~\cite{1709.08153}.

In this letter we investigate dynamical effects originating from the electron-electron interaction in the context of strong-field physics in solids, namely in the HHG emission spectra. 
To do so, we focus on a class of strongly correlated materials, the transition-metal oxides, which exhibit a wealth of interesting physical properties, covering superconductivity, magnetic materials, Mott or charge-transfer insulators 
\cite{zaanen_band_1985, imada_metal-insulator_1998, lee_doping_2006,keimer_quantum_2015}.
Those physical properties are governed by the localized and partially occupied $3d$ orbitals of the transition metal atoms. The electron-electron interaction in these orbitals is typically described by an effective on-site repulsion,  referred to as Hubbard $U$. %
Since $U$ plays a crucial role for the insulating behavior in charge-transfer and Mott insulators, these materials offer a testbed to challenge the general validity of the independent-particle approximation in strong-field physics.

Assuming that electrons remain independent during the interaction with an ultrafast driving field directly implies that any effective parameter describing the solid can be seen as constant during the interaction. To challenge this approximation, one therefore needs a theoretical framework capable of capturing the time evolution of $U$.
One possible approach to access this dynamics is to extend state-of-the-art methods, such as the constrained random-phase approximation (cRPA)~\cite{miyake_screened_2008,aichhorn_dynamical_2009,miyake_ab_2009,aichhorn_theoretical_2010,werner_satellites_2012},
to the time-dependent case. However, these methods have a prohibitive computational cost, and dynamical screening in Mott insulators without considering changes of $U$ has been studied in a model system only so far \cite{golez_dynamics_2015}.
In equilibrium, these methods have already highlighted the impact of dynamical-$U$ effects both on dynamical response functions and
thermodynamic phases in strongly correlated materials (see for instance Ref.~\cite{werner_satellites_2012}). 
In out-of-equilibrium settings such as strongly driven materials, the dynamical renormalization of $U$ is therefore expected to lead to similarly important consequences.

Here we use the recently proposed ACBN0 functional~\cite{agapito_reformulation_2015}, which can be seen as a pseudo-hybrid reformulation of the density-functional theory plus Hubbard $U$ (DFT+U) method.
This functional directly allows, by solving generalized Kohn-Sham equations, to compute the Hubbard $U$ and Hund's $J$ \textit{ab initio} and self-consistently, without the need of a supercell. This method has been recently extended by some of us to the real-time case\cite{Implementation_paper}, within the framework of time-dependent density-functional theory (TDDFT)~\cite{runge_density-functional_1984,van_leeuwen_mapping_1999}. TDDFT+U is a computationally efficient method to simulate the electronic response of systems driven out of equilibrium without relying on perturbation theory. It has been shown to yield accurate electronic band gaps, effective $U$, and surprisingly good linear absorption spectra of the charge-transfer insulators NiO and MnO \cite{Implementation_paper}. Motivated by these promising results, we propose here to employ this method to access the time evolution of $U$ of systems driven out of equilibrium.

Using this fully \textit{ab initio} TDDFT+U framework, we study for the first time HHG in bulk charge transfer insulators, taking nickel oxide as a prototypical material.
We show that in the nonperturbative regime required to obtain HHG, $U$ is strongly modified on the time-scale of the laser pulse.
We find that neglecting the dynamics of $U$ leads to a modified HHG spectrum, showing that dynamical effects originating from the electron-electron interaction can strongly affect the nonlinear properties of solids driven by strong laser fields.

All the calculations presented here were performed for bulk NiO, which is a type-II antiferromagnetic material below its his N\'eel temperature ($T_N=523$K\cite{cracknell_space_1969})\footnote{Below its N\'eel temperature, NiO exhibits a rhombohedral structure, which is obtained by contraction of the original cubic cell along one of the [111] directions.\cite{cracknell_space_1969} However, we have neglected the small distortions and considered NiO in its cubic rock-salt structure, which does not affect the result of calculated optical spectra. Calculations were performed using a lattice parameter of 4.1704~\AA\, a real-space spacing of $\Delta r=0.293$ Bohr, and a $28\times28\times28$ $\mathbf{k}$-point grid to sample the Brillouin zone. 
 We employ norm-conserving pseudo-potentials. For such few-cycle driver pulses, the HHG spectra from solids have been shown to be quite insensitive to the carrier-envelope phase (CEP), which is therefore taken to be zero here.}.
The driving field is taken along the [001] crystallographic direction in all the calculations. We consider a laser pulse of 25 fs duration (FWHM), with a sin-square envelope for the vector potential. The carrier wavelength $\lambda$ is 3000\,nm, corresponding to a carrier photon energy of 0.43\,eV. 
The time-dependent wavefunctions, current, and $U_{\mathrm{eff}}$ are computed by propagating generalized Kohn-Sham equations within TDDFT+U, as provided by the Octopus package.~\cite{andrade_real-space_2015}
We employed the PBE functional~\cite{perdew_generalized_1996} for describing the semilocal DFT part, and we computed the effective $U_{\mathrm{eff}}=U-J$ for the O $2p$ ($U^{2p}_{\mathrm{eff}}$) for one of the Ni $3d$ orbitals ($U^{3d}_{\mathrm{eff}}$), using localized atomic orbitals from the corresponding pseudopotentials~\cite{Implementation_paper}. Our ground-state values for $U$ and $J$ are consistent with other publications~\cite{Implementation_paper,agapito_reformulation_2015}.

\begin{figure}[t]
  \begin{center}
    \includegraphics[width=\columnwidth]{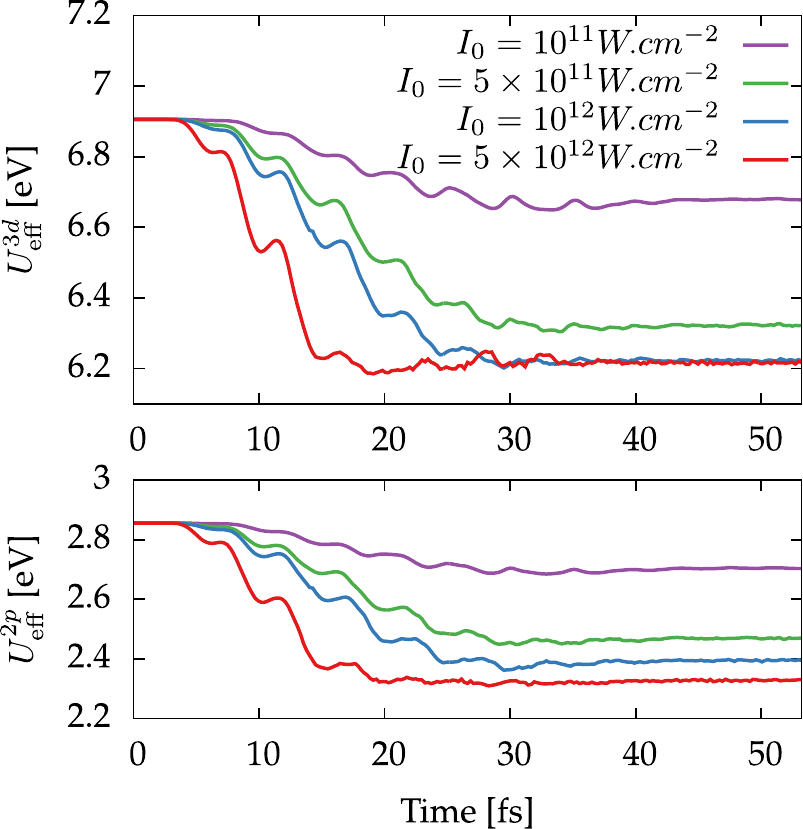}
    \caption{\label{fig:TD_U} Self-consistent dynamics of Hubbard $U$ for the Ni $3d$ orbitals (top panel), and the oxygen $2p$ orbitals (bottom panel) for pump intensities as indicated.}
  \end{center}
\end{figure}

\begin{figure}[t]
  \begin{center}
    \includegraphics[width=\columnwidth]{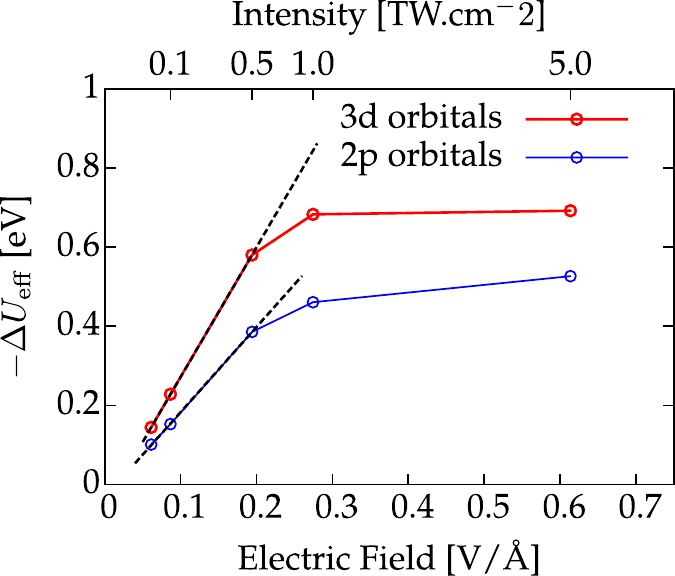}
    \caption{\label{fig:Scaling_TD_U} Calculated change of $U$ at the end of the laser pulse \textit{versus} the peak driving field strength, for the Ni $3d$ orbitals (red curve) and the O $2p$ orbitals (blue curve). The dashed lines indicate a linear scaling in electric field strength, as expected from linear response theory. The groundstate value of $U$ is 6.93 eV.}
  \end{center}
\end{figure}

From Fig.~\ref{fig:TD_U}  we find that the effective Hubbard $U$ is strongly modified by the applied laser, and that it decreases for both the O $2p$ and Ni $3d$ orbitals. Moreover, increasing the driving field intensity\footnote{Note that the intensity $I_0$ is defined here without the refractive index as $I_0=\frac{c\epsilon_0}{2}E^2$, where $E$ is the peak electric field strength, $c$ is the velocity of light in vacuum and $\epsilon_0$ is the vacuum permitivity.} leads to a stronger decrease of $U$. 
The reduction of $U$ is understood as follows. Dividing the Hilbert space into a localized subspace and the rest, as commonly done in cRPA~\cite{miyake_screened_2008,aichhorn_dynamical_2009,miyake_ab_2009,aichhorn_theoretical_2010}, it is clear that the screening, originating from the polarization associated with the rest of the system, increases as electrons are pumped from the localized subspace to the delocalized rest. This enhanced screening leads to a dynamical decrease of $U$.\\
It is worth noting that at higher intensities, the variation of $U$ is faster and then saturates to the same final value (see red lines in Fig.~\ref{fig:TD_U}). This saturation is expected, as the localized states only have a finite number of electrons to be excited, and therefore the decrease of $U$ must saturate as the intensity increases.
The final depletion of $U$, as taken from the long-time limit in Fig.~\ref{fig:TD_U}, is shown in Fig.~\ref{fig:Scaling_TD_U}. We observe that at moderate field strengths, the variation of $U$ is proportional to the electric field strength as expected from the linear response of the polarization, and then deviates from linear scaling, demonstrating a transition from linear to nonlinear reduction of $U$. We found that for moderate field strength, $-\Delta U_{\mathrm{eff}}^{3d} \approx 3.28 E$ and  $-\Delta U_{\mathrm{eff}}^{2p}\approx 2.15 E$, where $E$ is the electric field strength.
The dynamics of $U$ not only depends on the intensity of the driving pulse, but also depends on the shape and the length of the applied laser pulse.
From these results, it is clear that it is not possible to assume that the Hubbard $U$ remains constant during the interaction with the strong driving field, thus indicating in principle a breakdown of the widely assumed independent-particle or frozen band-structure approximation for strongly correlated materials. 

\begin{figure}[h!] 
  \begin{center}
    \includegraphics[width=\columnwidth]{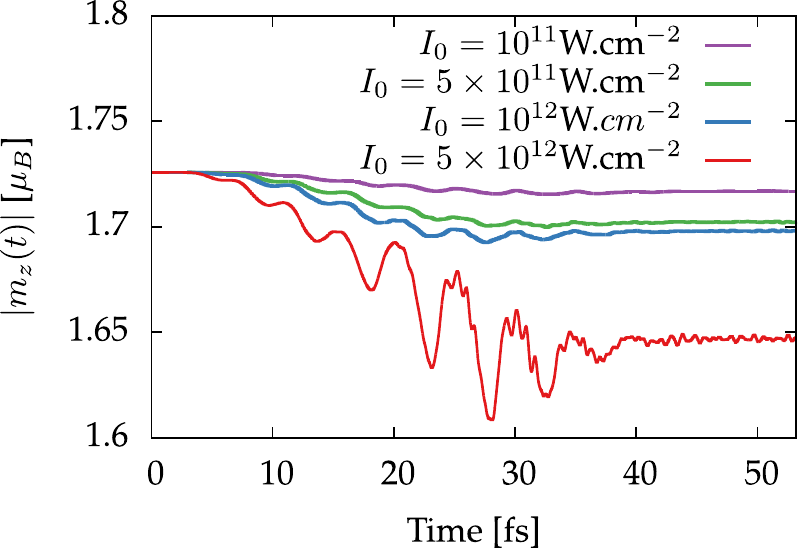}
    \caption{\label{fig:TD_mz} Time evolution of the magnetization integrated around the Ni atoms for different driving intensities. We only report the magnetization computed around one of the Ni atoms, as the two atoms are found to have exactly opposite magnetization, as expected for the antiferromagnetic phase considered here. The magnetization is calculated by averaging over a sphere of radius $1.97$ Bohr around the Ni atom.}
  \end{center}
\end{figure}

In order to confirm that the Hubbard $U$ reduction comes from the promotion of localized electrons to conduction bands, we also compute the magnetization of the system, spherically averaged around the Ni atoms. As shown in Fig.~\ref{fig:TD_mz}, the Ni magnetization decreases together with Hubbard $U$, thus confirming our physical interpretation.
Interestingly, we do not observe a saturation of the demagnetization of the Ni atoms, as opposed to the  final decrease of $U$ which is saturated for the corresponding laser intensities.

\begin{figure}[t]
  \begin{center}
    \includegraphics[width=0.95\columnwidth]{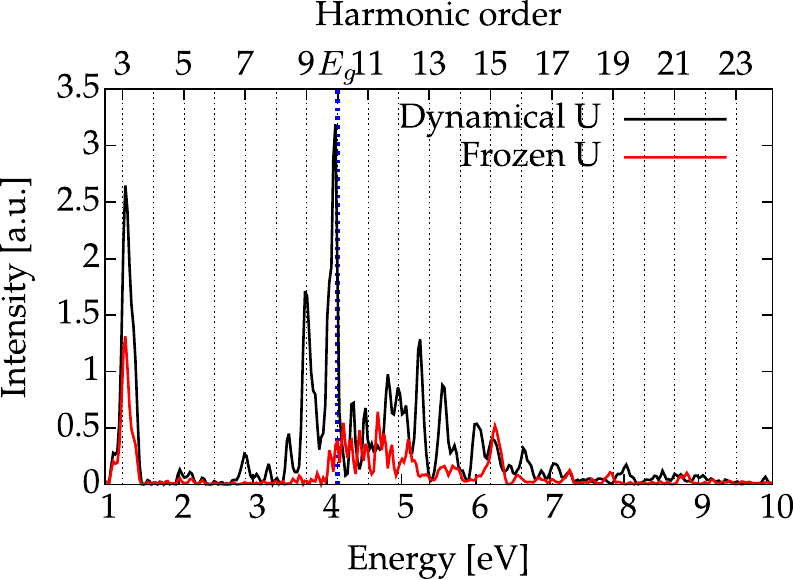}
    \caption{\label{fig:Comp_HHG} Effect of the time evolution of $U$ (from Fig.~\ref{fig:TD_U}) on the HHG spectrum of NiO.
    The HHG spectrum obtained from the full time evolution is shown in black, whereas the spectrum obtained for frozen $U$ is shown in red. The red vertical line indicates the calculated band gap of NiO. The intensity is taken here as $I_0=10^{12}$W\,cm$^{-2}$.}
  \end{center}
\end{figure}

To get more insight about the relevance of the dynamical Hubbard $U$ for describing the strong-field response of NiO, we compute its HHG spectrum with dynamical and static $U$ (Fig.~\ref{fig:Comp_HHG}). Form Fig.~\ref{fig:Comp_HHG} it is clear that neglecting the dynamics of $U$ in NiO leads to a strongly modified HHG spectrum (see Supplemental Information for different intensities and a comparison with PBE).
Indeed, freezing the Hubbard $U$ to its groundstate value of 6.93 eV implies that the band structure, and in particular the band gap of the material, remains constant during laser irradiation.
The light-induced reduction of $U$ (by about 10\%) implies that the band gap of the material becomes ``dinamically'' smaller during the laser pulse. As a direct consequence, the required excitation energy for Zener tunneling or multi-photon ionization decreases. Indeed this energy depends directly on the number of photons needed to reach the bandgap, thus reducing the band-gap leads to a stronger excitation of electrons and therefore to stronger harmonic emission.
Another important implication is that a dynamical $U$ introduces a novel dimension is the strong-field response of solids, i.e., time. Indeed, in usual semiconductors, changing the length of the driving pulse only results in a the change of the width of the harmonic peaks, as long as a pulse contains more than few optical cycles\cite{krausz_attosecond_2009,chang2016fundamentals}.
Effects such as the Stark effect~\cite{de_giovannini_monitoring_2016} or the renormalization of the bands due to the coupling to the laser pulse~\cite{hubener_creating_2017} are usually quite small for the laser parameters considered in HHG from solids. As the bandstructure of these solids remains unchanged on the time-scale of the optical pulse~\cite{tancogne-dejean_impact_2017}, and two consecutive non-overlapping pulses result in the same harmonic emission. However, for the case of NiO, a longer (shorter) pulse will lead to an more (less) important decrease of $U$, and a subsequent delayed pulse would feel a modify bandstructure, due to the reduction of $U$ by the first pulse, resulting in a modified harmonic emission.

In summary, we investigated the HHG spectra of the prototypical charge transfer insulator NiO driven by a strong below gap laser pulse excitation.
We showed that the strong laser intensity needed to generate high-order harmonics in NiO induces a significant change of Hubbard $U$ as a result of the increase of the screening from the photoexcited itinerant electrons.
Neglecting the change in $U$ strongly affects the solid HHG spectrum, demonstrating the measurable importance of the time dependence of $U$ in the nonlinear response of the material to a strong laser field.

Our results demonstrate that dynamical modification of the electronic parameters in correlated materials is indeed possible by purely electronic means without involving the crystal lattice. This should be contrasted with other dynamical perturbations, such as  for instance the concept of nonlinear phononics \cite{forst_nonlinear_2011, subedi_theory_2014}, in which direct excitations of optical phonons are used to induce such changes and trigger light-induced phase transitions. The time scales for phononic modifications of parameters are typically in the picosecond range, whereas here we showed that a much faster modification on femtosecond time scales is possible with our suggested mechanism, which might bear practical relevance for ultrafast switching processes. Conceptually, this faster mechanism may also allow us to disentangle electronic and phononic dynamics and also study cases in which such a decoupling is not possible, for instance when the Born-Oppenheimer approximation breaks down~\cite{tully_chemical_2000}. The proposed effect could be detected experimentally using pump-probe spectroscopy. One could measure band structure renormalizations in time-resolved angle-resolved photoemission spectroscopy, or one could detect the changes in THz conductivity or optical reflectivity in the time domain. 

From a methodological point of view, the developed TDDFT+U formalism allows for practical calculations from first principles of dynamically modulated $U$ and other relevant couplings in real materials.
In order to be efficient in controlling the dynamical properties of correlated materials, further studies should address the effect of the renormalization of $U$ due to other excitations such as for instance phonon dynamics, magnetic spin waves, heat fluctuations, and compare it with the present effect.
We expect that this work will pave the way for predicting and modeling interesting and potentially technologically relevant correlated materials for applications in ultrafast optoelectronics, magnetism, or spintronics.

\begin{acknowledgments}
A. R. acknowledges financial support from the European Research Council (ERC-2015-AdG-694097), Grupos Consolidados (IT578-13), and  European Union's H2020  program under GA no.676580 (NOMAD), and M. A. S. through the DFG Emmy Noether programme (SE 2558/2-1). 
We would like to thank O. D. M\"ucke and M. Altarelli for fruitful discussions.
\end{acknowledgments}

\bibliography{bibliography,HHG_TD_U}

\end{document}